\documentclass[10pt]{article}

\usepackage{amsmath}
\usepackage{amssymb}
\usepackage{indentfirst}

\newtheorem{theorem}{Theorem}
\newtheorem{lemma}{Lemma}

\renewcommand{\epsilon}{\varepsilon}
\renewcommand{\phi}{\varphi}
\newcommand{\Def}{\ensuremath{\stackrel{\rm def}=}} % define by equality
\newcommand{\ST}{\ensuremath{\quad\text{such that}\quad}}
\newcommand{\AND}{\ensuremath{\quad\text{and}\quad}}

\renewcommand{\H}{\ensuremath{{\cal H}}}
\renewcommand{\S}{\ensuremath{{\cal S}}}

\renewcommand{\L}{\ensuremath{{\cal L}}}
\newcommand{\X}{\ensuremath{{\cal X}}}
\newcommand{\Y}{\ensuremath{{\cal Y}}}
\renewcommand{\P}{\ensuremath{{\cal P}}}

\newcommand{\E}{\ensuremath{{\cal E}}}
\newcommand{\C}{\ensuremath{{\cal C}}}

\newcommand{\Hn}{\ensuremath{{\cal H}^{\otimes n}}}
\newcommand{\rhon}{\ensuremath{\rho^{\otimes n}}}
\newcommand{\sigman}{\ensuremath{\sigma^{\otimes n}}}

\newcommand{\rhoh}{\ensuremath{\hat{\rho}}}
\newcommand{\sigmah}{\ensuremath{\hat{\sigma}}}
\newcommand{\rhohn}{\ensuremath{\hat{\rho}^{\otimes n}}}
\newcommand{\sigmahn}{\ensuremath{\hat{\sigma}^{\otimes n}}}

\newcommand{\rhobar}{\ensuremath{\overline{\rho}}}
\newcommand{\Tr}{\ensuremath{\mbox{\rm Tr}}}
\newcommand{\Pe}{\ensuremath{\mbox{\rm Pe}}}

\newcommand{\proj}[1]{\ensuremath{\left\{ #1 \right\}}}
\newcommand{\oneover}[1]{\ensuremath{\frac{1}{#1}}}
\newcommand{\defset}[2]{\ensuremath{%
 \left\{#1\,\left|\,#2\right.\right\}
}}
\newcommand{\algenby}[1]{\ensuremath{[\![ #1 ]\!]}}
\newcommand{\pinching}[2]{\ensuremath{\,\E_{#1}\!\left(#2\right)}}
\newcommand{\Inner}[2]{\ensuremath{%
 \left\langle\!\langle\,#1\,,\,#2\,\right\rangle\!\rangle
}}
\newcommand{\bigzeroUR}{{\lower0.8ex\hbox{\Large 0}}}
\newcommand{\bigzeroDL}{{\raise0.8ex\hbox{\Large 0}}}
\def\QED{\mbox{\rule[0pt]{1.5ex}{1.5ex}}}

\def\endproof{\hspace*{\fill}~\QED\par\endtrivlist\unskip}
\def\keywords{\vspace{-.3em}
    \if@twocolumn
      \small\it Keywords\/\bf---$\!$%
    \else
      \begin{center}\small\bf Keywords\end{center}\quotation\small
    \fi}
\def\endkeywords{\vspace{0.6em}\par\if@twocolumn\else\endquotation\fi
    \normalsize\rm}
\def\appendix{\par
    \setcounter{section}{0}\setcounter{subsection}{0}
    \def\thesection{\Alph{section}} \section*{Appendix}
}
\def\appendices{\par
    \setcounter{section}{0}\setcounter{subsection}{0}
    \def\thesection{\Alph{section}} \section*{Appendices}
}
\begin{document}

\title{
A New Proof of the Channel Coding Theorem via Hypothesis Testing
in Quantum Information Theory
}

\author{
Tomohiro Ogawa
\thanks{
Department of Mathematical Informatics,
Graduate School of Information Science and Technology,
The University of Tokyo,
7--3--1 Hongo, Bunkyo-ku, Tokyo, 113--8656 Japan.
(e-mail: ogawa@sr3.t.u-tokyo.ac.jp)
}
\and
Hiroshi Nagaoka
\thanks{
Graduate School of Information Systems,
University of Electro-Communications,
1--5--1 Chofugaoka, Chofu, Tokyo 182--8585, Japan.
(e-mail: nagaoka@is.uec.ac.jp)
}
}

\date{}

\maketitle

\begin{abstract}
A new proof of the direct part of the quantum channel coding theorem
is shown based on a standpoint of quantum hypothesis testing.
A packing procedure of mutually noncommutative operators
is carried out to derive an upper bound on the error probability,
which is similar to Feinstein's lemma in classical channel coding.
The upper bound is used to show the proof of the direct part
along with a variant of Hiai-Petz's theorem in quantum hypothesis testing.
\end{abstract}

\begin{keywords}
Channel coding theorem,
hypothesis testing,
Hiai-Petz's theorem,
quantum relative entropy
\end{keywords}

%\vspace{-2ex}
\section{Introduction}
\label{section:introduction}

Let $\H$ be a Hilbert space which represents a physical system of
information carrier.
We suppose $\dim\H<\infty$ for mathematical simplicity.
Let $\L(\H)$ be the set of linear operators on $\H$
and define the totality of density operators on $\H$ by 
\begin{align}
\S(\H) \Def \defset{\rho\in\L(\H)}{\rho=\rho^*\ge 0, \Tr[\rho]=1}.
\end{align}
We will treat a quantum channel defined by a mapping
$x\in\X\longmapsto\rho_x\in\S(\H)$,
where $\X$ is a finite set of input alphabets
and each $\rho_x$ represents the quantum state of the output signal.

An encoding-decoding system of the message
over the $n$-th extension of the channel is described as follows.
Each message $k\in\{1,\dots,M_n\}$ is encoded to a codeword
$u^k=x_1^k \dots x_n^k$ in a codebook
$\C^{n}=\{u^1,\dots,u^{M_n} \}$ $\subseteq \X^{n}$ by an encoder,
where $\X^{n}$ is the $n$-th direct product of $\X$,
and the codeword $u^k$ is mapped to
$\rho_{u^k}=\rho_{x_1^k}\otimes\dots\otimes\rho_{x_n^k}\in\S(\Hn)$
through the channel.
The decoding process, which is called a decoder,
is represented by a set $X^{n}=\{X_1,X_2,\dots,X_{M_n}\}$
of nonnegative operators on $\Hn$
satisfying $\sum_{k=1}^{M_n} X_k \le I_n$,
implying that $X^{n}$ with $X_0\Def I_n-\sum_{k=1}^{M_n}X_k$
becomes a quantum measurement
on $\Hn$ taking its value in $\{0,1,\dots,M_n\}$.
A pair of encoding and decoding processes $(\C^{n},X^{n})$ is called a code
with cardinality $M_n$ or with transmission rate $R_n=\log M_n/n$.

The probability that the decoder outputs a message $k$
when a message $l$ is sent
is given by $\Tr[\rho_{u^l}X_k]$.
Thus, assuming that all messages
arise with the uniform probability, the average error probability of the
code $(\C^{n},X^{n})$ is given by
\begin{align}
\Pe\left(\C^{n},X^{n}\right)
\Def \oneover{M_n}\sum_{k=1}^{M_n} \left(1-\Tr[\rho_{u^k}X_k]\right).
\end{align}
Our interest lies in asymptotically achievable transmission rates
with arbitrarily small error.
The channel capacity is defined as the supremum of those values:
%%%%%%%%%%%%%%%%%%%%%%%%%%%%%%%%%%%%%%%%%%%%%%%%%%%%%%%%%%%%%%%%%%%%%%%%
\begin{align}
C \Def
\sup\biggl\{ R \,\biggm|\,
\exists\left\{ \left( \C^{n},X^{n} \right) \right\}_{n=1}^{\infty}
\ST
\lim_{n\rightarrow\infty}\Pe\left(\C^{n},X^{n}\right)=0
\AND
\liminf_{n\rightarrow\infty}\oneover{n}\log M_n \ge R
\biggr\}.
\end{align}

%%%%%%%%%%%%%%%%%%%%%%%%%%%%%%%%%%%%%%%%%%%%%%%%%%%%%%%%%%%%%%%%%%%%%%%%
In order to describe the quantum channel coding theorem,
let us introduce the quantum mutual information
\cite{Holevo-1973} as follows.
Let $\P(\X)$ be the totality of the probability distributions on $\X$,
and let $\sigma_p\Def\sum_{x\in\X} p(x) \rho_x$
be the mixture state by some $p\in\P(\X)$.
Then the quantum mutual information is defined by
\begin{align}
I(p) &\Def H(\sigma_p) - \sum_{x\in\X} p(x) H(\rho_x) \nonumber \\
&= \sum_{x\in\X} p(x) D(\rho_x\|\sigma_p),
\end{align}
where
$H(\rho) \Def -\Tr[\rho\log\rho]$
is the von Neumann entropy and
$D(\rho\|\sigma) \Def \Tr[\rho(\log\rho-\log\sigma)]$
is the quantum relative entropy.
One of the most significant theorems in quantum information theory
is the quantum channel coding theorem:
%%%%%%%%%%%%%%%%%%%%%%%%%%%%%%%%%%%%%%%%%%%%%%%%%%%%%%%%%%%%%%%%%%%%%%%%
\begin{align}
C=\max_{p\in\P(\X)}I(p).
\end{align}
%%%%%%%%%%%%%%%%%%%%%%%%%%%%%%%%%%%%%%%%%%%%%%%%%%%%%%%%%%%%%%%%%%%%%%%%

The theorem is composed of two inequalities:
the direct part ($\ge$) and the converse part ($\le$).
The direct part is given by showing the existence of a good code,
and was established in the middle of the 1990's
by Holevo \cite{Holevo-1998} and independently
by Schumacher and Westmoreland \cite{Schumacher-Westmoreland}
after the breakthrough by Hausladen {\it et al.} \cite{Hausladen-et-al}.
The classical counterpart of the method used there
is thought as a variant of the joint typical decoding \cite{Cover}
along with the random coding technique.
On the other hand, the converse part,
concerned with nonexistence of too good code,
goes back to 1970's works
by Holevo \cite{Holevo-1973} \cite{Holevo-1979}.
%The history of the quantum channel coding theory
%is described in the survey paper \cite{Holevo-survey}.
In 1999, another proof of the direct part
was also given by Winter \cite{Winter},
in which he developed the method of type
\cite{Wolfowitz} \cite{Csiszar-Korner} in the quantum setting
followed by a greedy construction of a good code.

In classical channel coding,
the most transparent proof of the direct part is thought to be
Feinstein's proof \cite{Feinstein} \cite{Blackwell-et-al}
(see also \cite{Ash}),
the essence of which is described below.
Let $w^n(y^n|x^n)$ be a channel matrix
transmitting an input sequence $x^n=x_1\dots x_n\in\X^n$
to an output sequence $y^n=y_1\dots y_n\in\Y^n$,
where $\X^n$ and $\Y^n$ are the $n$-th direct products of
the sets $\X$ and $\Y$ of input alphabets and output alphabets, respectively.
The channel is called stationary memoryless
if $w^n(y^n|x^n)=w(y_1|x_1)\dots w(y_n|x_n)$ holds
with a one-shot channel $w(y|x)\,(x\in\X,y\in\Y)$.
If the input $X^n=X_1\dots X_n$ is a random variable
taking its value in $\X^n$
subject to a probability distribution $p^n(x^n)\,(x^n\in\X^n)$,
then the output $Y^n=Y_1\dots Y_n$
and the pair $(X^n,Y^n)$ also become
random variables subject to
$q^n(y^n)=\sum_{x^n\in\X^n}p^n(x^n)w(y^n|x^n)$
and $p^n(x^n)w^n(y^n|x^n)$, respectively.
In the first part of Feinstein's proof, it is shown that
if we have
\begin{align}
\Pr\left\{ \oneover{n}\log\frac{w^n(Y^n|X^n)}{q^n(Y^n)} \le a \right\}
\longrightarrow 0 \quad (n\longrightarrow\infty)
\label{LLN}
\end{align}
for a real number $a$, then we can construct a reliable code
with transmission rate arbitrarily near $a$.
In the second part, assuming that
$p^n(x^n)$ is the independently and identically distributed (i.i.d.)
extension of a probability distribution $p(x)\,(x\in\X)$
and $w^n(y^n|x^n)$ is a stationary memoryless channel,
\eqref{LLN} is shown for $a<I(X;Y)$ by the law of large numbers,
where $I(X;Y)$ is the classical mutual information
for the random variables $(X,Y)$ subject to $p(x)w(y|x)$.
As for the first part, no assumption such as
the stationary memoryless property for the channel is needed,
which led to a general formula for the classical channel capacity
by Verd\'{u}-Han \cite{Verdu-Han} as one of the landmarks
in the information-spectrum method \cite{Han}.
As ways of providing the first part,
two different methods are known so far;
the random coding technique and the packing algorithm,
both of which give us an important insight into the construction of good codes.

In this paper
\footnote{
The prototype of the results was given in \cite{Ogawa-doctor}.
}
a new proof of the direct part for the stationary memoryless quantum channel
is shown based on a limiting theorem \cite{Ogawa-Hayashi}
in quantum hypothesis testing,
which is regarded as a variant of Hiai-Petz's theorem \cite{Hiai-Petz},
combined with a packing procedure for operators
following Winter \cite{Winter} to obtain a good code.
The limiting theorem in quantum hypothesis testing
is thought to be a substitute for the law of large numbers
used to show \eqref{LLN} in classical information theory.
The approach used here is regarded as
an attempt to develop the information-spectrum method
\cite{Han} in quantum channel coding,
which was followed by
Hayashi-Nagaoka \cite{Hayashi-Nagaoka} with
further developments using the random coding technique
to obtain a general formula for the channel capacity in the quantum setting.
It should be noted here, however,
any packing algorithm to derive the general formula is not
established yet.

%\vspace{-2ex}
\section{Relation with Hypothesis Testing}
\label{section:relation}

In the sequel, $\sigma_p=\sum_{x\in\X} p(x) \rho_x$
is written as $\sigma$ omitting the subscript $p$
when no confusion is likely to arise.
Let us define block diagonal matrices
\footnote{
The extensions of the density operators used here
were considered by \cite{Fujiwara-Nagaoka}
in an attempt to relate Hiai-Petz's theorem to channel coding.
},
%%%%%%%%%%%%%%%%%%%%%%%%%%%%%%%%%%%%%%%%%%%%%%%%%%%%%%%%%%%%
\begin{align}
\rhoh&\Def 
%%%%%%%%%%%%%%%%%%%%%%%%%%%%%%%%%%%%%%%%%%%%%%%%%%
\begin{pmatrix}
\ddots     &            & \bigzeroUR \\
           & p(x)\rho_x &            \\
\bigzeroDL &            & \ddots
\end{pmatrix}
%%%%%%%%%%%%%%%%%%%%%%%%%%%%%%%%%%%%%%%%%%%%%%%%%%
,&
\sigmah&\Def
%%%%%%%%%%%%%%%%%%%%%%%%%%%%%%%%%%%%%%%%%%%%%%%%%%
\begin{pmatrix}
\ddots     &              & \bigzeroUR   \\
           & p(x)\sigma &              \\
\bigzeroDL &              & \ddots
\end{pmatrix} ,
%%%%%%%%%%%%%%%%%%%%%%%%%%%%%%%%%%%%%%%%%%%%%%%%%%
\end{align}
%%%%%%%%%%%%%%%%%%%%%%%%%%%%%%%%%%%%%%%%%%%%%%%%%%%%%%%%%%%%
which are density operators in $\S\left(\bigoplus_{x\in\X}\H\right)$
and denoted as
\begin{align}
\rhoh&=\bigoplus_{x\in\X} p(x)\rho_x
,&
\sigmah&=\bigoplus_{x\in\X} p(x)\sigma ,
\end{align}
respectively.
It is important to note that the quantum mutual information
is nothing but the quantum relative entropy
between $\rhoh$ and $\sigmah$, i.e.,
$D(\rhoh\|\sigmah) = I(p)$.
In the same way the $n$-th tensor powers of $\rhoh$ and $\sigmah$
are given by the following block diagonal matrices
\begin{align}
\rhohn&=\bigoplus_{x^n\in\X^n} p^n(x^n)\rho_{x^n}
,&
\sigmahn&= \bigoplus_{x^n\in\X^n} p^n(x^n)\sigman
\end{align}
in $\S\left(\bigoplus_{x^n\in\X^n}\Hn\right)$, respectively,
where we used the following notations
for each $x^n=x_1x_2\dots x_n\in\X^n$
\begin{align}
p^n(x^n)&\Def p(x_1)p(x_2)\dots p(x_n), \\
\rho_{x^n}&\Def \rho_{x_1}\otimes\rho_{x_2}\otimes\dots\otimes\rho_{x_n}.
\end{align}
Here, let $\pinching{\sigmahn}{\rhohn}$ be the pinching
defined in Appendix~\ref{appendix:pinching}.
Applying Lemma~\ref{lemma:direct-sum-pinching}
in Appendix~\ref{appendix:pinching} inductively,
we can show that
\begin{align}
\pinching{\sigmahn}{\rhohn}
=\bigoplus_{x^n\in\X^n} p^n(x^n)
 \pinching{\sigman}{\rho_{x^n}}.
\label{extended-pinching}
\end{align}

In order to relate the above observation to the channel coding theorem,
let us introduce the quantum hypothesis testing problem here,
which is explained concisely in Appendix~\ref{appendix:Hiai-Petz},
and examine the error probability of a test
as follows.
Given a Hermitian operator $X= \sum_i x_i E_i$,
define the projection $\proj{ X > 0 }$ by
\begin{eqnarray}
\proj{ X > 0 } \Def \sum_{i:x_i > 0} E_i.
\label{Nagaoka-notation}
\end{eqnarray}
With the above notation, we define a test
for the hypotheses $\rhohn$ and $\sigmahn$ as
\begin{align}
\hat{S}_n(a)
&\Def \proj{\pinching{\sigmahn}{\rhohn} - e^{na}\sigmahn > 0}
\nonumber \\
&=\bigoplus_{x^n\in\X^n}
 \proj{\pinching{\sigman}{\rho_{x^n}} - e^{na}\sigman > 0},
\end{align}
where $a$ is a real parameter and
the last equality follows from \eqref{extended-pinching}.
The error probability of the first kind for the test
is written as follows by using the notation
$\rhobar_{x^n}\Def\pinching{\sigman}{\rho_{x^n}}$
\begin{align}
\hat{\alpha}_n\left(\hat{S}_n(a)\right)
&\Def \Tr\left[\rhohn\left(I_n-\hat{S}_n(a)\right)\right]
\nonumber \\
&= \sum_{x^n\in\X^n}p^n(x^n)
\Tr\left[\rhobar_{x^n}
\proj{\rhobar_{x^n} - e^{na}\sigman \le 0} \right],
\end{align}
which tends to zero exponentially if $a<D(\rhoh\|\sigmah)$
by Lemma~\ref{lemma:Ogawa-Hayashi}
in Appendix~\ref{appendix:Hiai-Petz} \cite{Ogawa-Hayashi}.
Thus we have shown the following lemma
as an analogue of \eqref{LLN} in quantum channel coding.
%%%%%%%%%%%%%%%%%%%%%%%%%%%%%%%%%%%%%%%%%%%%%%%%%%%%%%%%%%%%%%%%%%%%%%%%
\begin{lemma}
For $\forall a < I(p)$, we have
\begin{align}
\lim_{n\rightarrow\infty} \sum_{x^n\in\X^n}p^n(x^n)
\Tr\left[ \rhobar_{x^n}
\proj{\rhobar_{x^n} - e^{na}\sigman \le 0} \right]
=0.
\end{align}
\label{lemma:mutual-information-achiebable}
\end{lemma}
%%%%%%%%%%%%%%%%%%%%%%%%%%%%%%%%%%%%%%%%%%%%%%%%%%%%%%%%%%%%%%%%%%%%%%%%

\vspace{-4ex}
\section{Greedy Construction of a Code}
\label{section:packing}

In the previous section, we have shown the achievability of
the quantum mutual information in the sense of
Lemma~\ref{lemma:mutual-information-achiebable}.
Needless to say, this does not directly mean
the achievability concerned with
the error probability nor the transmission rate,
since the tests
$\proj{\rhobar_{x^n} - e^{na}\sigman > 0}$
for different $x^n$
do not commute each other and can not be realized simultaneously on $\Hn$.
In this section we will give a packing procedure of
the tests to meet the condition of
the quantum measurement on $\Hn$, which leads to the following lemma.
%%%%%%%%%%%%%%%%%%%%%%%%%%%%%%%%%%%%%%%%%%%%%%%%%%%%%%%%%%%%%%%%%%%%%%%%
\begin{lemma}
Let $x\in\X\longmapsto\rho_x\in\S(\H)$ be a quantum channel,
and let $\sigma\Def\sum_{x\in\X}p(x)\rho_x$ be the mixture state
by a probability distribution $p\in\P(\X)$.
Suppose that the following condition holds
for the pinching $\rhobar_x\Def\pinching{\sigma}{\rho_x}$
with some real numbers $\delta\ge 0$ and $c>0$:
\begin{align}
\sum_{x\in\X}p(x)\Tr\left[\rhobar_x\proj{\rhobar_x-c\,\sigma>0}\right]
 \ge 1-\delta.
\label{packing2}
\end{align}
Then, for any real numbers $\gamma>0$ and $\eta>0$,
there exist a codebook $\C=\{u_k\}_{k=1}^M\subseteq\X$
with cardinality $M$
and a decoder $X=\{X_k\}_{k=1}^M$ on $\H$ such that
\begin{align}
&\frac{\gamma}{\gamma+\delta}
\min\left\{\eta,1-\delta-\gamma-2\sqrt{\eta}\right\} \le \frac{M}{c},
\label{packing3} \\
&\oneover{M}\sum_{k=1}^M\left(1-\Tr[\rho_{u_k}X_k]\right)
\le \delta+\gamma+2\sqrt{\eta}.
\label{packing4}
\end{align}
\label{lemma:packing}
\end{lemma}
%%%%%%%%%%%%%%%%%%%%%%%%%%%%%%%%%%%%%%%%%%%%%%%%%%%%%%%%%%%%%%%%%%%%%%%%

\vspace{-2ex}
Before proceeding to the proof of Lemma~\ref{lemma:packing},
let us consider the $n$-th extension of the lemma
replacing the symbols above as follows:
\begin{align}
\X \longleftarrow \X^n,\quad
\rho_x \longleftarrow \rho_{x^n},\quad
p \longleftarrow p^n,\quad
\sigma \longleftarrow \sigman,\quad
c \longleftarrow e^{na},\quad
\gamma \longleftarrow \gamma_n,\quad
\eta \longleftarrow e^{-n\lambda},
\end{align}
where
$a$ is a real number and
$\lambda>0$ is an arbitrarily small number.
Letting $\rhobar_{x^n}=\pinching{\sigman}{\rho_{x^n}}$,
we can easily see that
\eqref{packing2} is satisfied
by setting the following $\delta_n(a)$ in place of $\delta$,
\begin{align}
\delta_n(a)
&\Def \sum_{x^n\in\X^n}p^n(x^n)\Tr\left[ \rhobar_{x^n}
 \proj{\rhobar_{x^n} - e^{na}\sigman \le 0} \right].
\end{align}
Therefore we obtain the following theorem using Lemma~\ref{lemma:packing}.

%%%%%%%%%%%%%%%%%%%%%%%%%%%%%%%%%%%%%%%%%%%%%%%%%%%%%%%%%%%%%%%%%%%%%%%%
\begin{theorem}
With the above notation,
there exist a codebook $\C^n=\{u^k\}_{k=1}^{M_n}\subseteq\X^n$
with cardinality $M_n$ and
a decoder $X^{n}=\{X_k\}_{k=1}^{M_n}$ on $\Hn$ such that
\begin{align}
&\frac{\gamma_n}{\gamma_n+\delta_n(a)}
\min\left\{
e^{-n\lambda},
1-\delta_n(a)-\gamma_n-2\sqrt{e^{-n\lambda}}
\right\}
\le e^{-na} M_n,
\\
& \Pe\left(\C^{n},X^{n}\right)
\le \delta_n(a) +\gamma_n+2\sqrt{e^{-n\lambda}}.
\end{align}
\end{theorem}
%%%%%%%%%%%%%%%%%%%%%%%%%%%%%%%%%%%%%%%%%%%%%%%%%%%%%%%%%%%%%%%%%%%%%%%%
For any $a<I(p)$, we have $\lim_{n\rightarrow\infty}\delta_n(a)= 0$ by
Lemma~\ref{lemma:mutual-information-achiebable},
and hence we can
choose $\gamma_n$ such that $\lim_{n\rightarrow\infty}\gamma_n=0$
and $\lim_{n\rightarrow\infty}\frac{\gamma_n}{\gamma_n+\delta_n(a)}$ $=1$.
Therefore the above theorem yields
\begin{align}
&\liminf_{n\rightarrow\infty}\oneover{n}\log M_n \ge a, \\
&\lim_{n\rightarrow\infty}\Pe\left(\C^{n},X^{n}\right)=0,
\end{align}
since $\lambda>0$ can be arbitrarily small.
Thus we have given a new proof of the direct part
of the quantum channel coding theorem
except that the proof of Lemma~\ref{lemma:packing} remains
to be shown.

\vspace{2ex}
%%%%%%%%%%%%%%%%%%%%%%%%%%%%%%%%%%%%%%%%%%%%%%%%%%%%%%%%%%%%%%%%%%%%%%%%
{\noindent{\it Proof of Lemma~\ref{lemma:packing}: }}
To begin with,
let us define a set of candidates for codewords by
\begin{align}
\X'\Def\defset{x\in\X}{
 \Tr[\rhobar_x\proj{\rhobar_x-c\,\sigma>0}]\ge 1-\delta-\gamma
}.
\end{align}
Then the probability of $\X'$ is bounded below as
\begin{align}
p(\X')\Def\sum_{x\in\X'}p(x) \ge \frac{\gamma}{\gamma+\delta},
\label{packing-proof0}
\end{align}
which is verified as follows:
\begin{align}
1-\delta
&\le \sum_{x\in\X} p(x)\Tr[\rhobar_x\proj{\rhobar_x-c\,\sigma>0}]
\nonumber \\
&= \sum_{x\in\X'} p(x)\Tr[\rhobar_x\proj{\rhobar_x-c\,\sigma>0}]
+ \sum_{x\in\X\backslash\X'} p(x)\Tr[\rhobar_x\proj{\rhobar_x-c\,\sigma>0}]
\nonumber \\
&\le p(\X') + (1-p(\X'))(1-\delta-\gamma).
\end{align}
Utilizing the normalization technique for operators developed by Winter
(see the proof of Theorem 10 in \cite{Winter}),
a codebook $\C=\{u_k\}_{k=1}^M\subseteq\X'$ and
a decoder $X=\{X_k\}_{k=1}^M$
are constructed by the following greedy algorithm,
along with the operators $S_k\in\L(\H)\,(k=0,1,\dots,M)$ for normalization.
%%%%%%%%%%%%%%%%%%%%%%%%%%%%%%%%%%%%%%%%%%%%%%%%%%%%%%%%%%%%%%%%%%%%%%%%
\begin{enumerate}
\item[(a)]
Let $S_0=0$.
\item[(b)]
Repeat the following procedures for $k=1,2,\dots$.
\begin{enumerate}
\item[(b-1)]
If there exists an $x\in\X'\backslash\{u_1,\dots,u_{k-1}\}$
such that $\Tr[\rhobar_{x}S_{k-1}]\le\eta$,
then choose such an $x$ arbitrarily and define
\begin{align*}
u_k&\Def x, \\
X_k&\Def \sqrt{I-S_{k-1}}\,\{\rhobar_{u_k}-c\,\sigma>0\}\sqrt{I-S_{k-1}}, \\
S_k&\Def S_{k-1}+X_k,
\end{align*}
else go to (c).
\item[(b-2)]
Let $k\longleftarrow k+1$ and go back to (b-1).
\end{enumerate}
\item[(c)]
Letting $M\longleftarrow k$,
$\C\Def\{u_k\}_{k=1}^M$
and $X\Def\{X_k\}_{k=1}^M$,
end the algorithm.
\end{enumerate}
%%%%%%%%%%%%%%%%%%%%%%%%%%%%%%%%%%%%%%%%%%%%%%%%%%%%%%%%%%%%%%%%%%%%%%%%
It should be noted here that each of $X_k$ and $S_k$
commutes with $\sigma$, and each $S_k$ satisfies the following inequalities
\begin{align}
0 \le S_k \le I,
\label{packing-proof1}
\end{align}
which follows from
\begin{align}
0 \le \proj{\rhobar_x-c\sigma>0} \le I
\end{align}
and induction.
Moreover it holds that for $\forall x\in\X'\backslash\C$
\begin{align}
\Tr[\rhobar_x S_M] > \eta,
\label{packing-proof2}
\end{align}
while for any codeword $u_k\in\C$ we have
$\Tr[\rhobar_{u_k}S_{k-1}]\le\eta$.
Thus, applying Winter's gentle measurement lemma
(Lemma~9 in \cite{Winter}, see Appendix~\ref{appendix:Winter}),
we obtain
\begin{align}
\left\|
\rhobar_{u_k} - \sqrt{I-S_{k-1}}\,\rhobar_{u_k}\sqrt{I-S_{k-1}}
\right\|_1 \le 2\sqrt{\eta}.
\end{align}
Therefore, using the property of the pinching \eqref{pinching-def},
we have
\begin{align}
\Tr[\rho_{u_k}X_k]
&= \Tr[\rhobar_{u_k}X_k]
\nonumber \\
&= \Tr\left[\sqrt{I-S_{k-1}}\,\rhobar_{u_k}\sqrt{I-S_{k-1}}
\proj{\rhobar_{u_k}-c\,\sigma>0}\right]
\nonumber \\
&= \Tr\Bigl[\rhobar_{u_k}\proj{\rhobar_{u_k}-c\,\sigma>0}\Bigr]
-\Tr\left[\left(
\rhobar_{u_k}-\sqrt{I-S_{k-1}}\,\rhobar_{u_k}\sqrt{I-S_{k-1}}
\right) \proj{\rhobar_{u_k}-c\,\sigma>0}
\right]
\nonumber \\
&\ge \Tr[\rhobar_{u_k}\proj{\rhobar_{u_k}-c\,\sigma>0}] - 2\sqrt{\eta}
\nonumber \\
&\ge 1-\delta-\gamma-2\sqrt{\eta},
\label{packing-proof3}
\end{align}
where the last inequality follows from $u_k\in\X'$,
and hence
\begin{align}
\Tr[\rhobar_{u_k} S_M]
&= \Tr\left[\rhobar_{u^k} \left(\sum_{k=1}^M X_k\right)\right]
\nonumber \\
&\ge \Tr[\rhobar_{u^k} X_k]
\nonumber \\
&\ge 1-\delta-\gamma-2\sqrt{\eta}.
\label{packing-proof4}
\end{align}
Now, in order to evaluate the cardinality of the code $M$,
we will estimate the lower and upper bound of $\Tr[\sigma S_M]$.
Using \eqref{packing-proof2} and \eqref{packing-proof4},
the lower bound is obtained as
\begin{align}
\Tr[\sigma S_M]
&= \sum_{x\in\X}p(x)\Tr[\rhobar_x S_M] \nonumber \\
&\ge \sum_{x\in\X'}p(x)\Tr[\rhobar_x S_M] \nonumber \\
&\ge \sum_{x\in\X'}p(x)
\min\left\{\eta,1-\delta-\gamma-2\sqrt{\eta}\right\} \nonumber \\
&\ge \frac{\gamma}{\gamma+\delta}
\min\left\{\eta,1-\delta-\gamma-2\sqrt{\eta}\right\},
\label{packing-proof5}
\end{align}
where the last inequality follows from \eqref{packing-proof0}.
On the other hand, $\Tr[\sigma S_M]$ is bounded above as
\begin{align}
\Tr[\sigma S_M]
&= \sum_{k=1}^M \Tr[\sigma X_k]
\nonumber \\
&= \sum_{k=1}^M
\Tr\left[
\sqrt{I-S_{k-1}}\,\sigma\sqrt{I-S_{k-1}}
\proj{\rhobar_{u_k}-c\,\sigma>0}
\right]
\nonumber \\
&\le \sum_{k=1}^M
\Tr\left[
\sigma \proj{\rhobar_{u_k}-c\,\sigma>0}
\right],
\label{packing-proof6}
\end{align}
since we have
$\sqrt{I-S_{k-1}}\,\sigma\sqrt{I-S_{k-1}} \le \sigma$,
which follows from \eqref{packing-proof1} and
the fact that each $\sqrt{I-S_{k-1}}$ commutes with $\sigma$.
Moreover,
\eqref{packing-proof6} is bounded above further as
\begin{align}
\Tr[\sigma S_M]
&\le \sum_{k=1}^M
\oneover{c} \Tr\left[
\rhobar_{u_k}\proj{\rhobar_{u_k}-c\,\sigma>0}
\right] \nonumber \\
&\le \frac{M}{c}
\label{packing-proof8}
\end{align}
by the definition of $\proj{\rhobar_{u_k}-c\,\sigma>0}$.
Now the assertion \eqref{packing3}
is shown by combining the lower bound \eqref{packing-proof5}
with the upper bound \eqref{packing-proof8},
and \eqref{packing4} follows from \eqref{packing-proof3}.
\endproof
%%%%%%%%%%%%%%%%%%%%%%%%%%%%%%%%%%%%%%%%%%%%%%%%%%%%%%%%%%%%%%%%%%%%%%%%

%\vspace{-2ex}
\section{Concluding Remarks}
\label{section:conclusion}
 
We have revisited the direct part of the quantum channel coding theorem
\cite{Holevo-1998} \cite{Schumacher-Westmoreland}
considering the relation with the hypothesis testing problem,
and an upper bound on the error probability
similar to Feinstein's lemma \cite{Feinstein}
has been obtained.
The approach used here is regarded as
an attempt to develop the information-spectrum method
\cite{Verdu-Han}
in quantum channel coding.

%\vspace{-2ex}
\appendices

%\vspace{-2ex}
\section{A property of the pinching}
\label{appendix:pinching}

We summarize a property of the pinching
related to a $*$-subalgebra in $\L(\H)$ for readers' convenience.
In the sequel, we denote the $*$-algebra generated by 
operators $\{I,A_1,A_2,\dots\}\subseteq\L(\H)$
as $\algenby{A_1,A_2,\dots}$.
Given a Hermitian operator $A\in\L(\H)$, 
let
\begin{align}
A=\sum_{i=1}^v a_iE_i
\label{spectral-decomposition}
\end{align}
be its spectral decomposition,
where each $E_i$ is the projection corresponding to an eigenvalue $a_i$
different from others and $v$ is the number of the eigenvalues.
Then we can easily see that
$\algenby{A} = \algenby{E_1,\dots,E_v}$ holds.
Actually, it is clear that we have
the inclusion $\algenby{A}\subseteq\algenby{E_1,\dots,E_v}$ from
(\ref{spectral-decomposition}).
On the other hand,
considering the following function for each $i$,
\begin{align}
f_i(x)\Def\prod_{j:j\neq i} \frac{x-a_j}{a_i-a_j},
\end{align}
we have
$E_i=f_i(A)$, which yields the converse inclusion
$\algenby{A}\supseteq\algenby{E_1,\dots,E_v}$.

Let us denote the commutant of $\algenby{A}$ as
\begin{align}
\algenby{A}'
&\Def \defset{B\in\L(H)}{\forall C\in \algenby{A}, \, BC=CB} \nonumber \\
&= \defset{B\in\L(H)}{BA=AB} \nonumber \\
&= \defset{B\in\L(H)}{BE_i=E_iB\,(i=1,\dots,v)} .
\end{align}
Then the pinching
$\E_A:\L(\H)\rightarrow\algenby{A}'\subseteq\L(\H)$ is defined
as the projection of an operator
to the $*$-subalgebra $\algenby{A}'$ so that
\begin{align}
\forall C\in \algenby{A}',\quad \Inner{B}{C}=\Inner{\E_A(B)}{C}
\label{pinching-def}
\end{align}
holds for $\forall B\in\L(\H)$, where
\begin{align}
\Inner{A}{B}\Def\Tr[A^*B]
\end{align}
is the Hilbert-Schmidt inner product.
Note that the pinching can be written explicitly as
\begin{align}
\E_A(B)=\sum_{i=1}^v E_i B E_i .
\end{align}

We used the following property of the pinching
to show \eqref{extended-pinching} in the section~\ref{section:relation}.
\begin{lemma}
Let $A,B,C,D\in\L(\H)$,
and define block diagonal matrices on $\H\oplus\H$ by
$X\Def A \oplus B$ and $Y\Def C \oplus D$
with the same notation as used in the section~\ref{section:relation}.
Then we have
\begin{align}
\E_Y(X)=\E_C(A) \oplus \E_D(B).
\label{direct-sum-pinching}
\end{align}
\label{lemma:direct-sum-pinching}
\end{lemma}
%%%%%%%%%%%%%%%%%%%%%%%%%%%%%%%%%%%%%%%%%%%%%%%%%%%%%%%%%%%%%%%%%%%%%%%%
\vspace{-4ex}
\begin{proof}
For any operator
$Z=\begin{pmatrix}
Z_{11} & Z_{12} \\
Z_{21} & Z_{22}
\end{pmatrix} \in \algenby{Y}'$,
$ZY=YZ$ yields $Z_{11}\in \algenby{C}'$
and $Z_{22}\in \algenby{D}'$ by a direct calculation.
Thus, we have
\begin{align}
\Tr[XZ]
&=\Tr[AZ_{11}]+\Tr[BZ_{22}] \nonumber \\
&=\Tr[\E_C(A)Z_{11}]+\Tr[\E_D(B)Z_{22}] \nonumber \\
&=\Tr\left[\,
\left(\E_C(A) \oplus \E_D(B)\right)
Z\,\right] .
\end{align}
Now the assertion has been proved,
since
$\E_C(A) \oplus \E_D(B)\in\algenby{Y}'$ is obvious
and $Z\in \algenby{Y}'$ is arbitrary.
\end{proof}

%\vspace{-2ex}
\section{Hiai-Petz's Theorem and its variants}
\label{appendix:Hiai-Petz}

In this appendix we summarize Hiai-Petz's theorem \cite{Hiai-Petz}
and its variants \cite{Ogawa-Hayashi}
in quantum hypothesis testing,
which is used to show Lemma~\ref{lemma:mutual-information-achiebable}.
Given $\rho$ and $\sigma$ in $\S(\H)$,
let us consider the hypothesis testing problem about
hypotheses $H_0 : \rhon\in\S(\Hn)$
and $H_1 : \sigman\in\S(\Hn)$.
The problem is to decide which hypothesis is true
based on a two-valued quantum measurement $\{X_0,X_1\}$ on $\Hn$,
where the subscripts $0$ and $1$ indicate the acceptance
of $H_0$ and $H_1$, respectively.
In the sequel, an operator $A_n\in\L(\Hn)$ satisfying inequalities
$0\le A_n\le I_n$ is called a test, since $A_n$ is 
identified with the measurement $\{A_n,\, I_n-A_n\}$. 
For a test $A_n$, the error probabilities of the first kind and 
the second kind are, respectively, defined by
\begin{align}
\alpha_n (A_n) &\Def \Tr[\rhon (I_n -A_n)],&
\beta_n (A_n) &\Def \Tr[\sigman A_n].
\end{align}

For $0<\forall\epsilon<1$, let us define
\begin{align}
\beta_n^*(\epsilon)\Def\min
\bigl\{ \beta_n(A_n) &\bigm| A_n:\text{test},\,
\alpha_n(A_n)\le\epsilon \bigr\},
\end{align}
and recall the quantum relative entropy:
\begin{align}
D(\rho\|\sigma)\Def\Tr[\rho(\log\rho-\log\sigma)].
\end{align}
Then we have the following theorem,
which is called the quantum Stein's lemma:
\begin{align}
\lim_{n\rightarrow\infty}\oneover{n}\log\beta_n^*(\epsilon)=-D(\rho\|\sigma).
\label{Stein}
\end{align}
The first proof of \eqref{Stein} was shown by two inequalities.
One is the direct part given by Hiai-Petz \cite{Hiai-Petz}:
\begin{align}
\limsup_{n\rightarrow\infty}\oneover{n}\log\beta_n^*(\epsilon)
\le -D(\rho\|\sigma),
\label{direct1}
\end{align}
and the other is the converse part
given by Ogawa-Nagaoka \cite{Ogawa-Nagaoka-2000}.

Preceding the direct part (\ref{direct1}),
Hiai-Petz \cite{Hiai-Petz} proved the following theorem
\begin{align}
D(\rho\|\sigma)
= \lim_{n\rightarrow\infty} \oneover{n}
\sup_{M_n} D_{M_n}\left(\rhon\big\|\sigman\right),
\label{Hiai-Petz-theorem}
\end{align}
where the supremum is taken over the set of quantum measurements on $\Hn$,
and $D_{M_n}\left(\rhon\big\|\sigman\right)$ is the classical relative entropy
(Kullback divergence) between the probability distributions
$\bigl\{\Tr[\rhon M_{n,i}]\bigr\}$
and  $\bigl\{\Tr[\sigman M_{n,i}]\bigr\}$.
Note that the monotonicity of the quantum relative entropy
\cite{Lindblad-CP} \cite{Uhlmann} yields
$D(\rho\|\sigma)\ge 1/n \cdot D_{M_n}\left(\rhon\big\|\sigman\right)$
for any measurement $M_n$, and in addition
there exists a measurement that attains the equality
if and only if $\rho$ and $\sigma$ mutually commute.
Hiai-Petz combined (\ref{Hiai-Petz-theorem}) with the classical hypothesis
testing problem to show the direct part  (\ref{direct1}).

In recent developments \cite{Hayashi-Stein} \cite{Ogawa-Hayashi},
we found direct proofs
of (\ref{direct1}) without using the achievability of
the information quantity (\ref{Hiai-Petz-theorem}).
Here, we will make use of a simple test
defined below with the notation \eqref{Nagaoka-notation}
and the pinching $\pinching{\sigman}{\rhon}$
(see Appendix~\ref{appendix:pinching}):
\begin{align}
\overline{S}_n(a)
\Def\proj{ \pinching{\sigman}{\rhon} - e^{na}\sigman > 0 },
\end{align}
which satisfies the following lemma
(see Theorem 2 in \cite{Ogawa-Hayashi}).

%%%%%%%%%%%%%%%%%%%%%%%%%%%%%%%%%%%%%%%%%%%%%%%%%%%%%%%%%%%%%%%%%%%%%%%%
\vspace{1ex}
\begin{lemma}[Ogawa-Hayashi]
For $0\le\forall s\le 1$, we have
\begin{align}
\alpha_n\left(\overline{S}_n(a)\right)
&\le (n+1)^{s\cdot\dim\H}\, e^{n[as-\psi(s)]},
\\
\beta_n\left(\overline{S}_n(a)\right)
&\le e^{-na} ,
\end{align}
where
\begin{align}
\psi(s)\Def -\log\Tr\left[
\rho\,\sigma^{\frac{s}{2}}\rho^{-s}\sigma^{\frac{s}{2}}
\right].
\end{align}
\label{lemma:Ogawa-Hayashi}
\end{lemma}
\vspace{-3ex}
%%%%%%%%%%%%%%%%%%%%%%%%%%%%%%%%%%%%%%%%%%%%%%%%%%%%%%%%%%%%%%%%%%%%%%%%
Observing that $\psi(0)=0$ and $\psi'(0) = D(\rho\|\sigma)$,
we can show that for $\forall a \,< D(\rho\|\sigma)$
\begin{align}
\lim_{n\rightarrow\infty}\alpha_n\left(\overline{S}_n(a)\right)&=0,&
\beta_n\left(\overline{S}_n(a)\right) &\le e^{-na}.
\end{align}
%with the exponential convergence of $\alpha_n\left(\overline{S}_n(a)\right)$.

%\vspace{-2ex}
\section{Winter's Lemma}
\label{appendix:Winter}

In this appendix
Winter's gentle measurement lemma
(Lemma~9 in \cite{Winter}) is explained.
The original proof of the lemma by Winter is based on a study
of the relation between the trace norm distance and the fidelity.
Here, a direct proof of the lemma is given for readers' convenience
accompanied by a little improvement of the constant in the upper bound.

%%%%%%%%%%%%%%%%%%%%%%%%%%%%%%%%%%%%%%%%%%%%%%%%%%%%%%%%%%%%%%%%%%%%%%%%
\begin{lemma}[Winter]
For $\forall\rho\in\S(\H)$ and $\forall X\in\L(\H)$
satisfying inequalities $0\le X\le I$,
we have
\begin{align}
\left\|
\rho - \sqrt{X}\rho\sqrt{X}
\right\|_1 \le 2 \sqrt{\Tr[\rho(I-X)]}.
\end{align}
\end{lemma}
\vspace{-1ex}
%%%%%%%%%%%%%%%%%%%%%%%%%%%%%%%%%%%%%%%%%%%%%%%%%%%%%%%%%%%%%%%%%%%%%%%%
\begin{proof}
A direct calculation yields
\begin{align}
\left\| \rho - \sqrt{X}\rho\sqrt{X} \right\|_1
&= \left\|
\left(I-\sqrt{X}+\sqrt{X}\right)\rho
-\sqrt{X}\rho\sqrt{X}
\right\|_1
\nonumber \\
&\le \left\| \left(I-\sqrt{X}\right)\rho \right\|_1
+\left\| \sqrt{X}\rho\left(I-\sqrt{X}\right) \right\|_1
\label{Winter1} \\
&= \Tr\left|\left(I-\sqrt{X}\right)\sqrt{\rho}\cdot\sqrt{\rho}\right|
+\Tr\left|\sqrt{X}\sqrt{\rho}\cdot
\sqrt{\rho}\left(I-\sqrt{X}\right)\right|
\nonumber \\
&\le
\sqrt{
\Tr\left[\rho\left(I-\sqrt{X}\right)^2\right]
\cdot \Tr[\rho]
}
 +\sqrt{
\Tr\left[\rho X \right]
\cdot \Tr\left[\rho\left(I-\sqrt{X}\right)^2\right]
}
\label{Winter2} \\
&\le 2 \sqrt{\Tr[\rho(I-X)]},
\label{Winter3}
\end{align}
where \eqref{Winter1} follows from the triangle inequality,
\eqref{Winter2} follows from the Cauchy-Schwartz inequality for operators,
and we used $\left(1-\sqrt{x}\right)^2\le(1-x)$
for $0\le x\le 1$ and $\Tr[\rho X]\le 1$ in \eqref{Winter3}.
\end{proof}
%%%%%%%%%%%%%%%%%%%%%%%%%%%%%%%%%%%%%%%%%%%%%%%%%%%%%%%%%%%%%%%%%%%%%%%%

%\vspace{-2ex}
\section*{Acknowledgment}

This research was partially supported by
the Ministry of Education, Culture, Sports, Science, and Technology
Grant-in-Aid for Encouragement of Young Scientists,
13750058, 2001.

%\bibliography{channel}

\end{document}